\documentclass[amsmath,amssymb,11pt]{revtex4-1}

\usepackage{latexsym}
\usepackage{amsmath}
\usepackage{amsfonts}
\usepackage{amssymb}
\usepackage{amsbsy}
\usepackage{bm}
\usepackage{color}
\usepackage{graphicx}
\usepackage{longtable}
\usepackage{setspace}
\usepackage{color}
\usepackage{hyperref}

\usepackage{float}

\providecommand\bcdot{\boldsymbol{\cdot}}
\providecommand\bnab{\mbox{\boldmath $\nabla$}}

\usepackage [left, running] {lineno}
 
\begin{document}

\begin{titlepage}
\title{The propagation of active-passive interfaces in bacterial swarms}
\author{Alison E. Patteson$^{1,+,*}$, Arvind Gopinath$^{2,3}$ and  Paulo E. Arratia$^{1,*}$}
\affiliation{$^{1}$Department of Mechanical Engineering \& Applied Mechanics,
University of Pennsylvania, Philadelphia, PA 19104.\\
$^{2}$Department of Bioengineering, University of California Merced, CA 95340. \\
$^{3}$Health Sciences Research Institute, University of California Merced, CA 95340\\
$^{+}$Present Address - Physics Department, Syracuse University NY 13244.}

\begin{abstract}
{Propagating interfaces are ubiquitous in nature, underlying instabilities and pattern formation in biology and material science. Physical principles governing interface growth are well understood in passive settings; however, our understanding of interfaces in active systems is still in its infancy. Here, we study the evolution of an active-passive interface using a model active matter system, bacterial swarms. We use ultra-violet light exposure to create compact domains of passive bacteria within {\it Serratia marcescens} swarms, thereby creating interfaces separating motile and immotile cells. Post-exposure, the boundary re-shapes and erodes due to self-emergent collective flows. We demonstrate that the active-passive boundary acts as a diffuse interface with mechanical properties set by the flow. Intriguingly, interfacial velocity couples to local swarm speed and interface curvature, suggesting that an active analogue to classic Gibbs-Thomson-Stefan conditions controls boundary propagation. Our results generalize interface theories to mixing and segregation in active systems with collective flows.}
\end{abstract}
\maketitle
\end{titlepage}

Bacteria live and move in an extraordinarily wide range of habitats and can quickly respond to the presence of other cells 
and physical boundaries in their environment. For instance, bacteria swim independently in fluids, but when 
transferred to surfaces display a collective behavior known as swarming \cite{Harshey2003,Kearns2010}.  Swarming occurs in many gram-negative and gram-positive species and corresponds to a hyper-flagellated elongated phenotype \cite{Kearns2010,Alberti1990,Harshey1994}. Swarming cells self-organize into rapid collective motions  that allow for quick colonization of new environmental niches \cite{Steager2008,Darnton2010,Harshey2003,Living2008}. Swarming is co-regulated with virulence determinants, inversely-regulated with sessile biofilm formation, and associated with enhanced antibiotic resistance \cite{Living2008,Kim2003,Butler2010}. More broadly, the collective motion of self-propelling (active) particles \cite{Marchetti2013,Patteson2016_2} is observed in bacterial infections \cite{Stoodley2004}, embryogenesis \cite{Mayor2016}, and wound healing~\cite{Meyer2008} and is an important feature of both prokaryotic \cite{BenJacob1998,Koch2011} and eukaryotic \cite{Ladoux2013,Vicsek2014} systems.

Ecological niches are typically a heterogeneous mix of cells, and internal boundaries can form separating cells of two difference types. Bacterial swarms coexist symbiotically with other microbes -- assisting in the transport of fungal spores 
\cite{Ingham2011} and other bacterial species \cite{Fink2015} --  or they compete at sharp boundaries 
\cite{science2008,budding2009}. Boundaries also emerge in cultures of the same bacteria due to chemotaxis and cell death 
\cite{Berg1995,Cates2010} or the presence of extracellular polymers \cite{Schwarz2012,Levine2014,Cicuta2014}, both of 
which can induce a swimming-speed dependent phase separation \cite{Cates2010,Schwarz2012}. Segregation of 
active particles is not unique to biological settings, arising in synthetic systems such as phoretic particles 
\cite{Palacci2013,Brady2015}.

In passive bi-phasic systems \cite{Bray2003,Langer,Moroney2014} such as melting ice-water mixtures and solidifying alloys, 
properties of internal boundaries (such as interface shape and speed) depend on the surface tension, interfacial energies, 
and externally imposed flows. In active systems, particle motion can couple to the presence of boundaries which can lead to 
anomalous properties in mechanical pressure \cite{Takatori2014,Cates2015} and impact collective flows 
\cite{Goldstein2014,Gompper2016}. However, despite the ubiquity of boundaries in living and life-like materials, boundary 
stability and motion remain largely unexplored in active non-equilibrium matter. Identifying boundary conditions that link boundary structure and active motion will help elucidate a minimal description of actively-driven boundaries.

Here, we focus on the propagation of an interface separating mobile and immobile bacteria in swarming \emph{Serratia marcescens}. \emph{Serratia} are a rod-shaped, gram-negative, opportunistic pathogen of the Enterobacteriaceae family \cite{Alberti1990}. We use high-intensity ultra-violet (UV) light to selectively paralyze and passivate cells in large compact domains within the swarm (Methods). The passive domain and active swarm interact at the interphase boundary where self-emergent, vortical flows develop. The interphase is spontaneously reshaped and eroded as passivated bacteria are dislodged from their neighbors and convected by nearby collective flows. Intriguingly, the active-passive boundary behaves as a propagating, diffuse elastic interface with speeds that correlate with local interface curvature and the intensity of the active bacterial flow. Our results suggest the existence of an active analog to the classic Gibbs-Thomson-Stefan boundary conditions of passive multiphase systems.

\begin{figure*}[]
\includegraphics[width=.95\columnwidth]{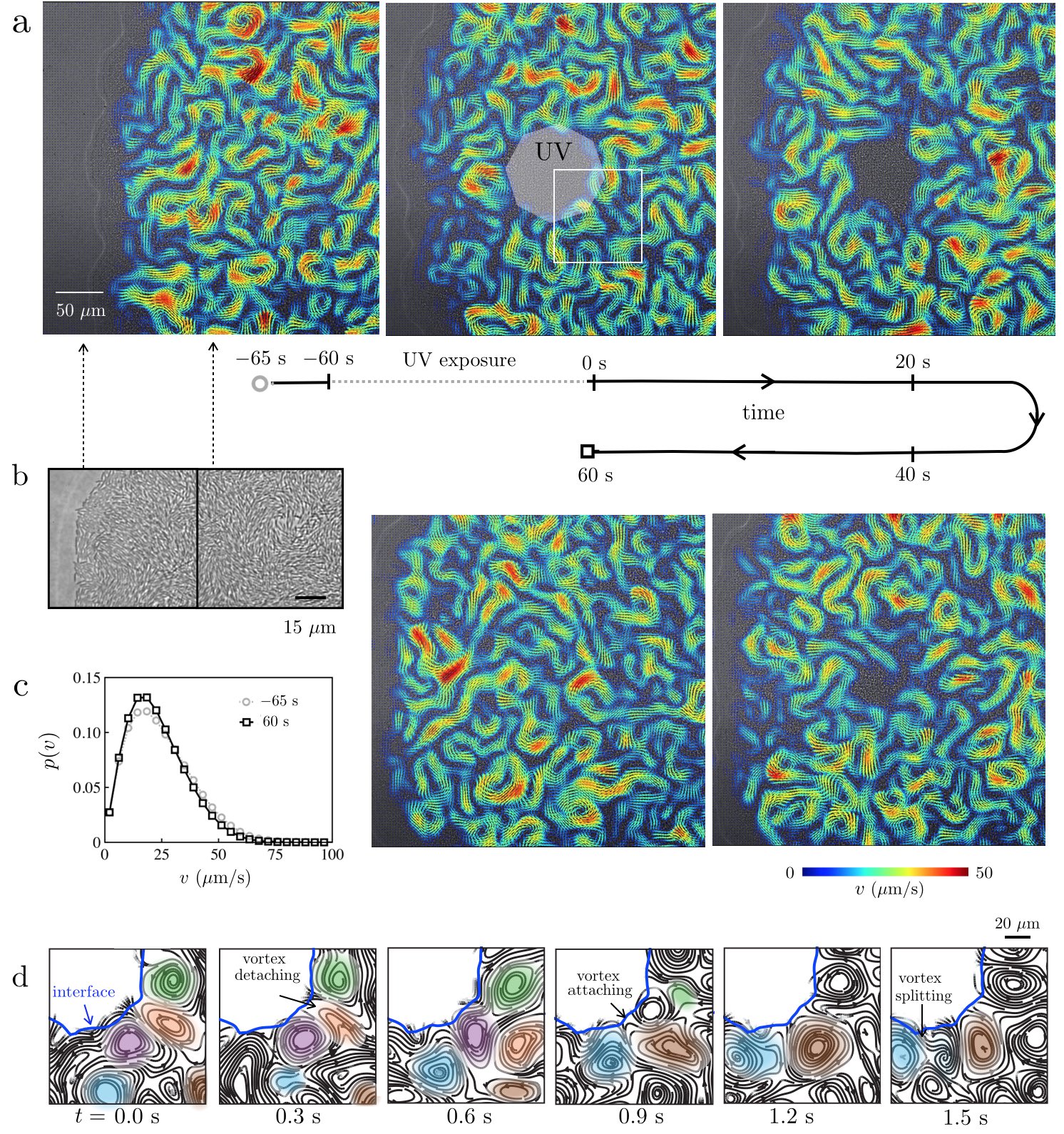}
\caption{{\bf {Creation and dissolution of an active-passive interphase boundary in a bacterial swarm}} 
{\bf (a)} Snapshot of the expanding edge of a swarm of \emph{Serratia marcescens} on agar; the colony front exhibits a precursor fluid film ($\sim$ 5 $\mu$m; white curve), moving from right to left. The swarm shows long-range collective flows, with strong velocity fields (PIV; overlaid color). A large domain of passive, immobile bacteria is created by exposing a region of the swarm to high intensity ultraviolet (UV) light (highlighted octagon). An interphase boundary forms between the passivated and active bacteria. When the light source is switched off ($t=0$ s), the active unexposed bacteria deform the interphase boundary and penetrate the passive region. Over time, active bacteria convect immobile bacteria away from the passive domain, causing the boundary to erode and propagate inward. The swarm dissolves the passive phase in 60 s, with interface speeds greater than that of expanding colony edge. {\bf (b)} The swarm edge (close-up) features densely packed cells with local polarity and nematic order. {\bf (c)} The swarm's collective motion recovers after dissolution as shown by the probability distribution of bacterial speeds $p(v)$, before and after exposure. {\bf (d)} A montage of the flow streamlines - from the highlighted box in (a) - reveals the motion of vortices (labeled by color) at the interface (blue line). Vortices starting in the bulk can collide and attach to the interface (labeled brown vortex for example); some vortices at the surface detach and move away (green, orange). Others fade away (purple) or split (dark blue splits from light blue vortex in right tile).} 
\label{F1}
\end{figure*}

Figure \ref{F1}a shows snapshots of a {\em Serratia marcescens} swarm before and after its exposure to UV light. The swarm is grown on an agar substrates (Methods \ref{M1}) and is pictured moving right to left at a speed of approximately 1 $\mu$m/min; the colony edge is marked by a clear precursor fluid film (white edge in image). Close examination reveals (Fig. \ref{F1}b) densely-packed rod-shaped cells with local orientational order resembling a nematic liquid crystal. The individual cells \cite{Alberti1990} are {1} $\mu$m in diameter and {5-30} $\mu$m in length, and the collective swarm edge is estimated to be approximately a monolayer thick based on previous investigations \cite{Darnton2010}. In its initial state, the swarm is highly motile and exhibits long-range collective flows (SI-Movie 1). We use particle image velocimetry (PIV, Methods \ref{M2}, SI-$\S$II) to exact the bacterial velocity field $ {\bf v}({\bf r}, t)$ (overlaid in color). We find that the fastest moving cells are approximately 100-400 $\mu$m from the leading edge of the colony, where the average local speed is 28 $\mu$m/s (Fig. \ref{F1}c). The collective flow has a correlation length of approximately 20 $\mu$m and characteristic time of 0.25 seconds (SI-$\S$II, SI-Fig.~1).

When exposed to sufficiently high intensity light (Methods \ref{M3}, SI-I), a portion of the highly-motile swarm is quenched and becomes immobile (Fig. \ref{F1}a, SI-Movie 2). We use standard microscope optics to focus the light  from a mercury lamp source to the swarm, selectively damaging and blocking cell motility [Methods \ref{M3}, SI-I]. This generates domains of passive bacteria, the shape of which can be controlled through an aperture. We focus on two aperture geometries: a half-plane aperture [H] and an octagonal-shaped [O] aperture.

Figures \ref{F1} illustrates the swarm's response to UV light using the [O] aperture. Bacteria inside the exposed region (highlighted octagon) stop moving and are eventually trapped. This is seen in SI-Movie~2: particles that serve as tracers slow down and eventually stop moving as they are trapped amongst the passive bacteria. Bacteria outside the exposure remain motile, forming vortices and jets along the boundary of the trapped cells (SI-Movie~3).

The collective flow of bacteria shapes and erodes the passive domain. When the light source is switched off, the active unexposed bacteria penetrate the passive region, dislodge, and thereafter convect passive bacteria away from the interface (SI Movie 4). The interface propagates radially inward albeit asymmetrically. The swarm's collective motion recovers to its pre-exposure state (Fig. \ref{F1}c).

Flow streamlines (Fig. \ref{F1}{d}) near the interphase boundary show the rich complex behaviors of individual vortices. We observe vortices convecting in from the bulk, interacting with each other, and lingering at the surface (highlighted brown vortex, for example). Similar features occur when the exposure is made using the [H] aperture (SI-Fig. 2, SI-Movie 5); a notable difference for the [H] aperture is that the absence of imposed corners and the approximate translational invariance along the exposed edge results in active-passive boundary propagating without large scale curvatures.

Bi-phasic passive systems involving boundary erosion such as melting of ice or alloy solidification usually feature a diffuse interfacial region where the two phases mix. 
We hypothesized that, similarly, the boundary between the passive (immobile) and active (swarming) bacteria may be treated as an diffuse interface possessing an intrinsic time dependent thickness, $w$. We also hypothesized that a suitable (mesoscale level) continuous phase-field order parameters could be used to define an interface with physically relevant properties. Thus we defined two order parameters $\phi$ (Methods \ref{M4}, SI-III): (i) $\phi$ - based on intensity fluctuations between images of the swarm, which are related to bacterial density fluctuations, and (ii) $\phi_v$ - based on the bacterial velocity fields. 

\begin{figure*}[t]
\includegraphics[width=.9\columnwidth]{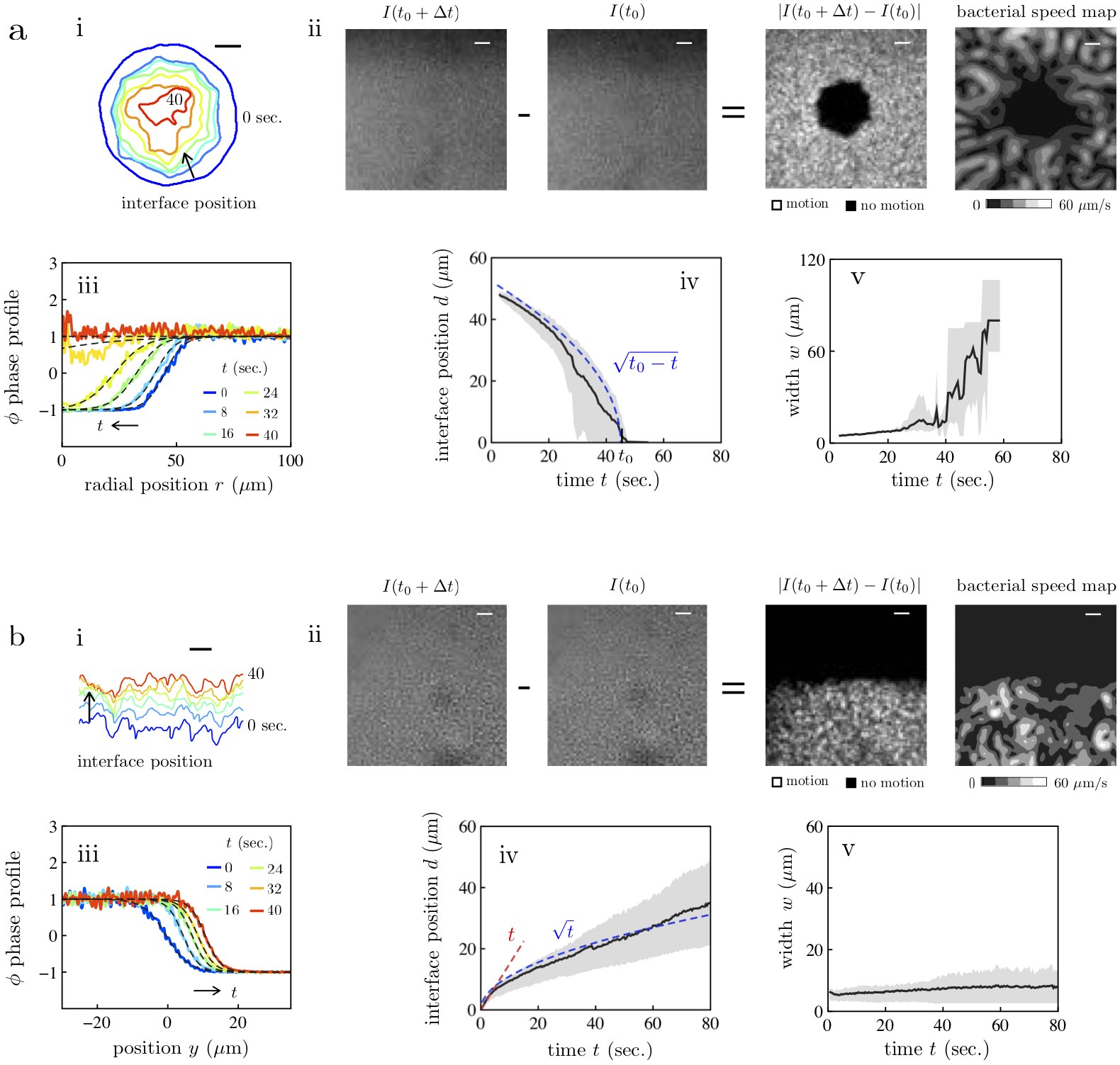}
\caption{{\bf (Color Online) Growth and structure of the active-passive interface in the bacterial swarm} Active-passive domain boundaries are designed with different initial shapes by varying the geometry of the aperture, such as an octagon [O] (a) or half-space [H] (b). ({\bf a-i \& b-i}) The interface position changes shape as it moves over time.  ({\bf a-ii \& b-ii}) The interface position is identified through order parameters $\phi$ based on fluctuations in image intensity ($\Delta t = 0.1$ s) and $\phi_v$, the bacterial speed (Methods \ref{M4}). The interface positions in (ai) and (bi) correspond to $\phi^*=0$. For the [O] aperture, the interface roughens as the domain shrinks in size; for [H], the roughness appears to stabilize. ({\bf a-iii \& b-iii}) We find that averaged one-dimensional profiles of the $\phi$-fields possess a smooth transitions between the active ($\phi=1$) and passive ($\phi=-1$) phases. Fits of the data (dashed lines, Eqn. 6 \& 7 Methods) yield the mean location $d$ and width $w$ of the interface. ({\bf a-iv and b-iv}) Parameters $d$ and $w$ are determined for multiple experiments ($N = 4$ per condition) and averaged together as a function of time (min to max variation in grey). For the [O] aperture, the mean interface position $d(t)$ (black line, measured radially from the center of the domain) decreases over time and follows the scaling law $d \sim \sqrt{(t_0-t)}$ with $t_{0} \approx 40$ s. For the [H] aperture, the interface position $d$ initially follows $d\sim t$ (red line) and then transitions to  faster than $d\sim \sqrt{t}$ (blue line) at $t \approx 2$ sec. Width $w$ ({\bf a-v \& b-v}) ranges from 4 and 10 $\mu$m and varies little over time - except for the case of the octagon aperture at long times ($t >$ 40 s) as the passive domain dissolves entirely. Scale bar, 50 $\mu$m.} 
\label{F2}
\end{figure*}

Each order parameter defines a phase field $\phi(r,t)$ that quantifies the local motion throughout the swarm as shown in Fig. \ref{F2}: the active phase corresponds to values of +1 and the passive phase, -1. A transitional interfacial region exists between the two phases; we define the mathematical interface as the locus of points ${\bf r}_{s}$ satisfying $\phi({\bf r}_{s}, t)=0$ (Fig. \ref{F2}a-i \& b-i). We find that the results obtained from using the order parameters $\phi$ and $\phi_v$ yield essentially the same description of the active-passive interface as seen in the phase field maps in Fig. \ref{F2}a-ii \& b-ii; here, we present results primarily from $\phi$ based on intensity fluctuations (see SI-Fig. 4 for $\phi_v$ results).

To examine the boundary thickness, we spatially-average the phase field into one-dimensional phase profiles, $\phi^{*}$ (Fig. \ref{F2}a-iii \& b-iii, Methods \ref{M5}). Physically, we intuit that the one-dimensional, spatially-averaged phase profiles correspond to a propagating, broadening, and coarsening interface, whose analytical form corresponds to a hyperbolic tangent. Indeed, our phase profiles are well captured using fits to hyperbolic tangent functions (Fig. \ref{F2}a-iii \& b-iii, Methods \ref{M5}), which yield the mean interface position $d$ and interface thickness $w$ over time.

The evolution of the mean interface position $d$ depends on the shape of the aperture (Fig. \ref{F2}a-iv \& b-iv). For the (O) aperture, $d(t)$ provides an area averaged radius of the passive domain that decreases over time, eventually reducing to zero as the whole domain erodes away; in this case, we find $d \sim \sqrt{t_0-t}$ (Fig. \ref{F2}a-iv), with the time at which the passive phase disappears, $t_0 \approx 40$ s. For the (H) aperture (Fig. \ref{F2}b-iv), the mean interface position propagates faster than $d\sim \sqrt{t}$ for the duration of the experiment.  

The interface thickness $w$ ranges from 4 to 10 $\mu$m ($\sim$ bacterial length) over most of the dissolution process (Fig. \ref{F2}a-v \& b-v). The most significant deviation from this trend is seen for the [O] aperture at relatively long times (Fig. \ref{F2}a-v): when the interface width is approximately the radius of the passive phase ($t \approx 40$ s), both $w$ and $d$ exhibit large fluctuations and the thickness increases dramatically. {Our results show that} the interface width is a weak function of time (Fig. \ref{F2}a-v \& b-v), consistent with a quasi-steady propagating interface. We do observe isolated bacteria - singly or in very small pockets - entering into the passive region {at distances more than the interface thickness} (SI Movie 4). {However, these events are rare,} resulting in the averaging procedure yielding a phase-field based interface without overhangs. We interpret $w$ as a correlation length characterizing the gradient in the density of active bacteria.

The active-passive phase fields based on intensity fluctuations ($\phi$) and bacterial velocity field ($\phi_v$) yield qualitatively similar results (SI-III). For instance, the interface positions are the same (SI-Fig.~4). We hypothesize that $w_v$ is greater than $w$ (SI-Fig. 4) because velocity fields vary over vortex (20~$\mu$m) length scales whereas intensity fluctuations vary over bacterial ($5$~$\mu$m) scales.

\begin{figure*}[t]
\includegraphics[width=.85\columnwidth]{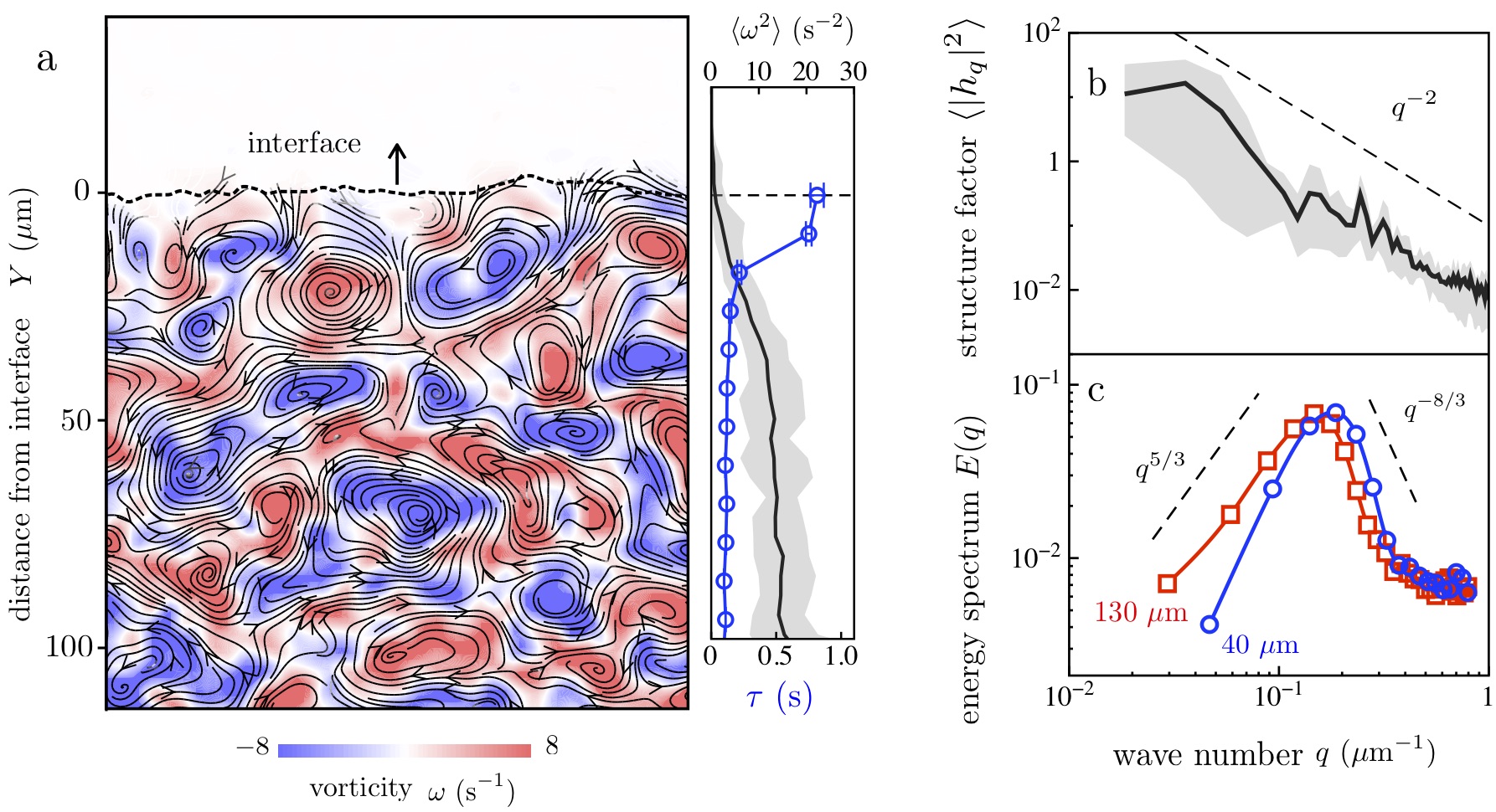}
\caption{{\bf (Color Online) Interface structure is coupled to collective flow of bacteria} {\bf (a)} A snapshot of the bacterial vorticity field, $t = 4$ s after exposure with the half-space (H) aperture.  {Right inset,} the strength of the vorticity field, characterized by $\langle \omega^2  \rangle$, increases as one moves away from the interface boundary; while the flow decay timescale $\tau$ decreases (grey variation and error bars from temporal fluctuations). {\bf(b)} The overall decay of the interface static structure factor scales as $|\Delta h_q|^2\sim q^{-2}$, yielding an interfacial stiffness $\kappa=$ 0.7 $\mu$m$^{-2}$ ($N=4$; grey representing min to max variation). {\bf(c)} The energy spectrum of the flow $E(q)$ (Equation 9) is non-monotonic with a peak that shifts to higher values of $q$ close to the interface ($Y \approx 40$ $\mu$m) compared with the bulk active phase ($Y \approx 130$ $\mu$m). The peaks in $E(q)$ coincides with peaks in $|\Delta h_q^2|$.}  
\label{F3}
\end{figure*}

Next, we examine how the bacterial flow interacts with the interface. Figure \ref{F3}a highlights two robust features of the flow near the interface: (i) a gradient in flow perpendicularly across the interface and (ii) an array of size-varying vortical flows. We find that the flow varies in both strength and dynamics. As shown in \ref{F3}a (inset), the square of vorticity $\langle \omega^2 \rangle$ increases as one moves away from the interface; while the flow decay time $\tau$ decreases (Methods \ref{M6}), indicating an increase in vortical flow lifetimes, a result similar to recent simulations of active and passive sphere mixtures \cite{Gompper2016}. Overall, our data suggests a coupling between the interface and flow: the interface plays a stabilizing role on the collective flow while the gradient in bacterial vorticity marks a flux of momentum and energy from the swarming bacteria to the interface, energy that can be used to erode and reform the surface. Consistent with this picture, the interface undergoes displacements through interactions with many fluctuating vortical flows (Methods \ref{M6}).

To understand the boundary-flow coupling, we calculate the (quasi-static) structure factor of the interface (Fig. \ref{F3}b) and the two-dimensional energy spectrum of the flow (Fig. \ref{F3}c). We focus here on experiments using the [H] aperture, which are not limited by finite time dissolution effects. The static structure factor $\langle |\Delta h_q|^2\rangle$ (Methods \ref{M6}) decays with wavenumber $q$, punctuated by peaks at $q = 0.15$ $\mu$m$^{-1}$ and 0.22 $\mu$m$^{-1}$ (Fig. \ref{F3}b). For $q > 0.4$ $\mu$m$^{-1}$, the overall decay scales as $\langle {|\Delta h_q|}^2\rangle\sim {q}^{-2}$.  To correlate these length scales with the energetic features of the flow,  we calculate the two-dimensional time-averaged energy spectra $E(q)$ from the bacteria velocity field (Methods \ref{M6}) for different distances from the interface $Y$. As shown in Fig. \ref{F3}c, $E(q)$ is non-monotonic with $q$ and depends on $Y$. This non-monotonic behavior is a unique characteristic of active fluids and is attributed to the injection of energy at the level of the bacteria \cite{Wensink2012}. Compared to dense suspensions of swimming bacteria in microfluidic devices \cite{Wensink2012}, our measured spectra exhibit similar scalings with $q$ as $q^{5/3}$ and $q^{-8/3}$  (Fig. \ref{F3}c) with a peak  centered at scales of about 1-2 vortex sizes. The interface also impacts the flow spectra by shifting the peak to higher $q$, indicating a shift to smaller vortices near the interface (Fig. \ref{F3}c). 

For passive fluid-fluid interfaces fluctuating due to white noise, equipartition of energy requires that  $\langle {|\Delta h_q|}^2\rangle=({\kappa A q^2})^{-1}$ with $\kappa$ being the stiffness, and $A$ the interface area. Equipartition does not hold in our system. Comparing $E(q)$ to $\langle |\Delta h_q|^2\rangle$ (Fig. \ref{F3}b \& c), we find that the peak in the $E(q)$ (between 0.15-0.2 $\mu$m$^{-1}$; length scale, 15-20 $\mu$m) correlates with peaks in $ \langle |\Delta h_q|^2\rangle$ - a signature that the interface is forced at length scales of the flow structures (streamers and vortices). Still, since the scaling $\langle |\Delta h_q|^2\rangle \sim q^{-2}$ provides a good description over a range of length scales (Fig.~\ref{F3}b), we use this form to fit the data (Methods \ref{M7}) and extract an interfacial stiffness $\kappa \approx 0.7 $ $\mu$m$^{-2}$. This value is much smaller than the interfacial stiffness of water in air (1.7$\times 10^7$ $\mu$m$^{-2}$) but similar to phase-separated systems involving colloids \cite{Hernandez2009,Gasser2001,Aarts2004} (0.1-20 $\mu$m$^{-2}$). This stiffness is to be interpreted as an effective value of the diffuse interface and inherently accommodates variations in passive bacteria alignment.

Here, the bacterial flow injects energy into the surface at various wave-numbers and frequencies. One way to qualify the energetic features of the flow is by invoking an effective temperature $T_{\mathrm{eff}}$. In the absence of flow, $T_{\mathrm{eff}}$ can be estimated by following tracers that sample the system for sufficiently long times \cite{Patteson2016}. Our swarm features collective flows; $T_{\mathrm{eff}}$ therefore is ambiguous. Nonetheless, using 2 $\mu$m polystyrene spheres as tracer particles (Methods \ref{M7}, SI-Fig. 10), we estimate $T_{\mathrm{eff}}\approx 2.2 \times 10^{5}$ K, which yields an apparent surface tension $\kappa k_B T_{{\mathrm{eff}}} \sim 10$ pN/$\mu$m.

In the paradigmatic example - melting of ice - the interface between ice and water propagates with speeds controlled by the temperature field at the interface and heat flux across the interface. For a stationary ice-water boundary, the Gibbs-Thomson relationship provides the relationship between the interface curvature ${\mathcal{C}}$ and the temperature (equivalently, chemical potential) while the Stefan boundary condition constrains the flux of heat across the interface. For slowly propagating ice-water interfaces, the Gibbs-Thomson relationship requires extension; a linear relationship can be obtained between the local curvature, local interface velocity and temperature (or chemical potential) \cite{Langer,Moroney2014}.

\begin{figure*}[]
 \includegraphics[width=.9\columnwidth]{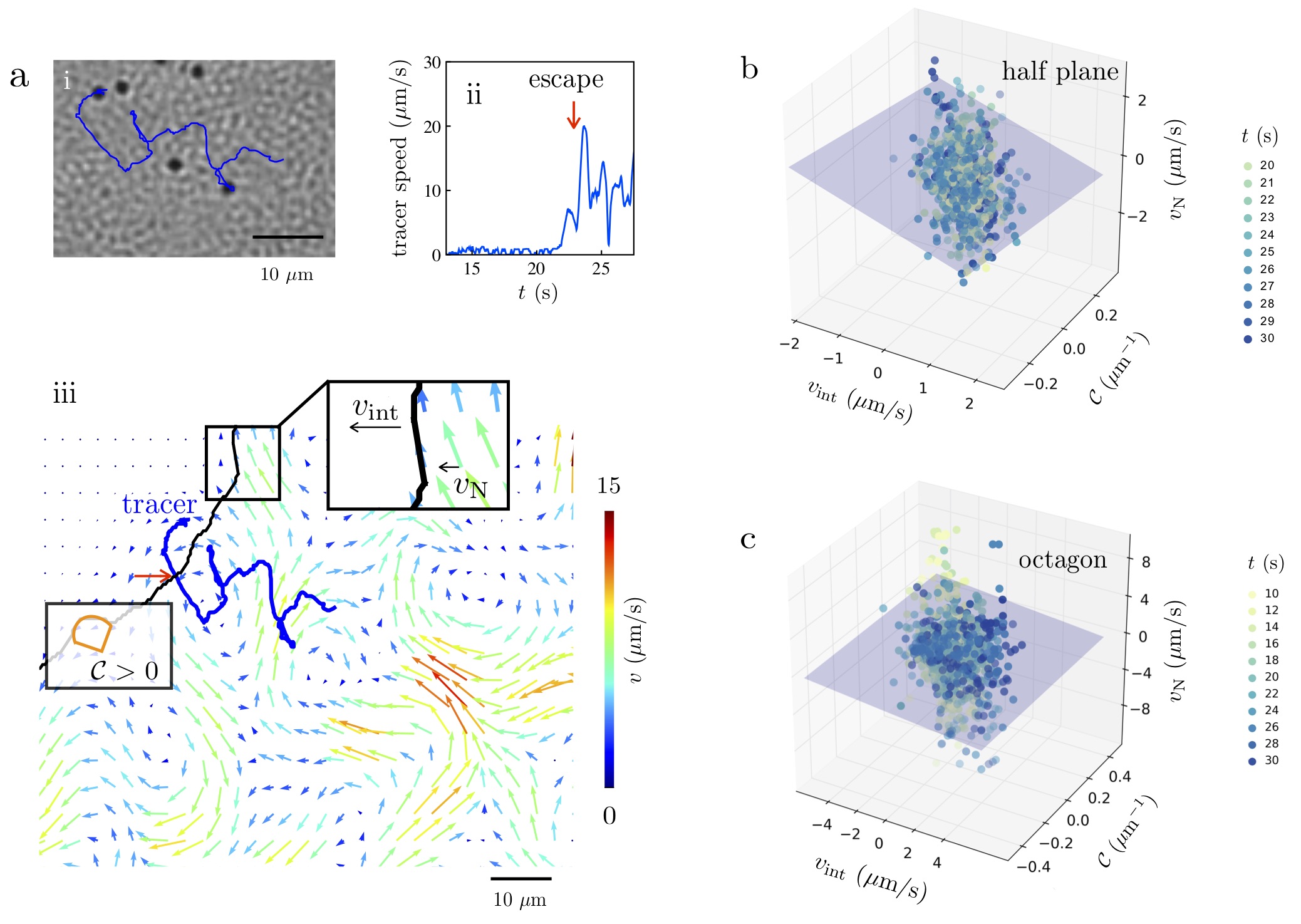}
\caption{{\bf (Color Online) Evolution of the active-passive interface follows an active analogue to classical Gibbs-Thomson-Stefan conditions} ({\bf a}) Particle motion at the interface. (i) The trajectories of 2 $\mu$m polystyrene tracer particles, reveal how the swarm extracts particles from the passive domain. First, particles in the passive phase caged-in by their neighbors do not move. Second, when the diffuse interphase is roughly a velocity correlation length $O(w_{v})$ away, the trapped particle starts to fluctuate, as seen in the time trace of its speed (b). Third, agitations induced by the active particles  dislodge the jammed tracer from the cage of neighboring passive particles. For the tracer in (iii), the flow field moves the particle tangentially along the interface (SI Movie 6). Finally, the particle escapes the passive domain as normal streaming flows between adjacent vortices pull the particle away, primarily perpendicularly from the interface. This escape correlates with a rapid increase in particle speed; once inside the swarm, the speed of the tracer fluctuates (red arrow in ii and iii). The schematic in (iii) defines the sign of the curvature and normal bacterial velocity. {\bf (b,c)} Local interface velocity $v_{\textrm{int}}$ correlates with the interface curvature $\mathcal{C}$ and the normal component of the collective bacterial velocity $v_{\textrm{N}}$ at the interface. Data for $v_{\textrm{int}}$, $\mathcal{C}$, and ($v_{\textrm{N}}$), gathered over time, collapses unto a plane for both {\bf (b)}  half-plane [H]  and the {\bf (c)} octagonal [O] apertures. This suggests an active analog of the classical Gibbs-Thomson- Stefan boundary conditions exists.} 
\label{F4}
\end{figure*}

To test for analogous boundary conditions for active-passive interfaces, we visualized how active flows extracted and convected passive particles at the boundary (Fig.\ref{F4}a). Based on these observations (see Fig. 4(a), Methods \ref{M8}) we propose a simple linear relation that captures the main ingredients of the erosion process relating the interface kinematics to geometry and intensity of activity,
\begin{equation}
v_{\textrm{int}}= a_{1} + a_{2}\mathcal{C}+ a_{3} {\mathcal{X}}
\label{AGT}
\end{equation}
with the variable ${\mathcal{X}}$ standing in for the swarm activity. We chose bacterial speed $|v|$, $v_{\textrm{N}}$ (normal component of the bacterial velocity), $v_{\mathrm{T}}$ (tangential component), and vorticity~$\omega$ as possible stand-ins for ${\mathcal{X}}$ and evaluated the fidelity of the fit of Eqn. \ref{AGT} to the experimental data (Methods \ref{M8}, SI-$\S$V).  We gather these data at multiple points along the interface for varying times throughout the experiment. The data is combined with local measures of the interface velocity and interface curvature and cast into three-dimensional scatter plots (SI-$\S$ V,SI-Fig. 9). The best collapse of the data to a plane ($p < 0.05$) was obtained for $v_\textrm{N}$, as shown in Fig. \ref{F4}b $\&$ \ref{F4}c (SI-$\S$V, Methods \ref{M8}). 
The values of the coefficients are dependent on the aperture geometry. 
For the half-plane [H] aperture, we find
$a_1 = 0.4$ $\mu$m/s, 
$a_2 = -2.3$  $\mu$m$^2$/s and 
$a_3 = -2.0$. For the octagonal [O] aperture, the presence of a mean curvature shifts the plane, which we find for $v_{\textrm{N}}$ to be given by $a_1=0.02$ $\mu$m/s, $a_2=-1.12$ $\mu$m$^2$/s, and $a_3=-0.7$.
At a qualitative level, the scatter plots in Fig. \ref{F4} demonstrate three-way coupling between the interface velocity, bacterial flow, and curvature in a manner reminiscent of the classical Gibbs-Thomson-Stefan boundary conditions in passive systems. The interface velocity is negatively correlated with curvature $\mathcal{C}$. Regions of negative curvature move into the passive phase as passive bacteria are extracted from the surface.

We conclude that dissolution of passive domains within active swarms is a spontaneous self-organized process governed by short-range interactions at the single bacterial level and long-range non-local effects due to collective flow. Three features define the active-passive interface during the erosion process. First, the interphase region acts as a broadening diffuse interface with a well defined position and thickness. Second, the interface is stiff with elastic constants dependent on flow intensity; stronger flows are expected to lower the effective stiffness, allowing for faster erosion.  Third, the interface structure is tightly coupled to the statistics of the collective swarming flows. The interphase boundary stabilizes vortices that form parallel to the surface, enabling sustained erosion. The net result is that the interface erosion follows a relation bearing similarities to the classic Gibbs-Thomson-Stefan boundary conditions. These results generalize classical interface theories that couple mechanics, thermodynamics and kinetics to active living and synthetic systems with collective flows.

Interfaces separating cells of differing motility is a common motif in natural habitats, including biofilms, where swimmers, spores, and other non-motile phenotypes segregate into different domains \cite{Kolter2011}. Starting with light-induced boundaries in a model system of \emph{Serratia marcescens} swarms, we identified an active version of Gibbs-Thomson and Stefan boundary conditions. We anticipate that these physical mechanisms underlie segregation and pattern formation in active biological and synthetic settings. Natural next steps would be to explore how extracellular polymers - which are implicated in both single cell swimming \cite{Berg1979,Patteson2015} and collective biofilm expansion \cite{Gloag2013} - impact boundary dynamics in microbial environments.

\section*{Methods}

%Preparation of bacterial swarms and imaging
\subsection{Preparation of bacterial swarms and imaging}
 \label{M1}
\noindent{}We use swarming \emph{Serratia marcescens} cultured on on agar substrates. The agar is prepared by dissolving 1 w\% Bacto Tryptone, 0.5 w\% yeast extract, 0.5 w\% NaCl, and 0.6 w\% Bacto Agar in ddH$_2$O. This is melted and poured into petri dishes after addition of 2 w\% of glucose solution (25 w\%). 
Once the agar cools and solidifies, \emph{Serratia marcescens} (strain ATCC 274, Manassas, VA) from frozen glycerol stocks are inoculated on the agar plates and incubated at 34$^{\mathrm{o}}$ C. Bacterial colonies form at the inoculation sites and grow outward on the agar substrate, away from the inoculation site. Experiments are performed 12-16 hours after inoculation; the bacteria are imaged with the free surface facing down with an inverted microscope.
The bacteria were imaged with an inverted Nikon microscope Eclipse Ti-U with a Nikon 20x (NA = 0.45) and Nikon 60x (NA = 0.7) objective. Images were gathered at either 60 or 125 frames per second with a Photron Fastcam SA1.1 camera.

\subsection{Bacterial velocity fields from PIV}

\label{M2}
\noindent{}Bacterial velocity fields, $ {\bf v}({\bf r}, t)$, were extracted at 3 $\mu$m intervals from videos using particle image velocimetry (PIV, SI-$\S$II). The velocity fields reveal complex transient collective flows at the expanding edge of the colony. Selecting regions of the swarm spanning 100 to 500 $\mu$m from the colony edge - where the swarming behavior appears strongest - we calculated the spatial and temporal velocity correlation functions,  $C_r(\Delta r)$  and  $C_t (\Delta t)$ using
\begin{equation}
C_r(\Delta r) = \Big \langle \frac{{\bf v}(r_0)\cdot {\bf v}(r_0+\Delta r)}{|{\bf v}(r_0)|^2} \Big \rangle, 
\end{equation}
and
\begin{equation}
C_t(\Delta t) = \Big \langle \frac{{\bf v}(t_0)\cdot {\bf v}(t_0+\Delta t)}{|{\bf v}(t_0)|^2} \Big \rangle.
\end{equation}
Here, the brackets denote and average over reference times $t_{0}$ and reference positions $r_{0}$. Pre-exposure,  the swarming behavior does not appear to be biased in any particular direction (the swarming speeds are much larger than the speed of the advancing colony front); hence, scalar function defined by (1) and (2) suffice to characterize the flow.  From these measurements, we extract a characteristic vortex size of 20 $\mu$m and vortex (residence or lifetime) timescale of 0.24 seconds (SI-$\S$II, SI-Fig. 1).

\subsection{Exposure to high intensity ultraviolet light}
\label{M3}
\noindent{}{To create a passive phase of immobile bacteria, we used standard fluorescence microscope optics to expose cells to high intensity light. The light source is an unfiltered mercury lamp, which has a wide-spectrum that includes a significant amount of ultra-violet (UV) light (100-400 nm). The response of bacteria to light depends on the wavelength, intensity, and duration of the light \cite{Taylor1975} and is affected by photosensitizers \cite{Taylor1975,Berg1984}. High intensity light $\geq$ 200 mW/cm$^2$ at wavelengths 390-530 nm can lead to increased tumbling, slow swimming, and eventual irreversible paralysis \cite{Taylor1975,Berg1984}. It is hypothesized that paralysis may be due to flagellar motor damage caused by photosensitizing flavins and dyes that are present in growth media such as LB Broth. 
Experiments \cite{Yeow2013} with \emph{Bacillus subtilis} also suggest that photosensitizers can be used to reversibly or irreversibly affect collective motility. Cell death occurs from photodynamic action and UV light-induced DNA damage. 

Consistent with previous experiments \cite{Taylor1975,Berg1984,Yeow2013}, we found that for swarming \emph{Serratia marcescens},  passivation was not immediate and sometimes reversible; yet for sufficiently large light intensities and exposure times they were rendered permanently passivated. We selected an exposure time of 60 s and light intensity of $I = 370$ $\mu$W (measured at 535 nm) to reproducibly immobilize \emph{Serratia marcescens} (SI-I).}

\subsection{{Phase-field order parameters identify interface}}
\label{M4}
\noindent{}We use two scalar order parameters, $\phi$ and $\phi_{v}$, based on a phase-field description of the interface region between the active (motile) and passive (non-motile) domains to define, identify, locate and characterize the interface boundary (see also SI-$\S$III).  

{(I) From intensity fluctuations:} The first scalar order parameter, $\phi$ is computed from intensity fluctuations, $\Delta I$, defined by the difference in image intensities taken at 0.1 second intervals of the bacterial swarm (SI-$\S$III-1).  Treating the intensity fluctuation as a measure of the fluctuations in the bacteria density, we define the order parameter at location ${\bf r}$ by
\begin{equation}
\phi({\bf r}, t)=\frac{2|\Delta I({\bf r}, t)|-|\Delta I_A (t)|-|\Delta I_P(t)|}{|\Delta I_A(t)|-|\Delta I_P(t)|}.
\end{equation}
where $|\Delta I_A (t)|$ and $|\Delta I_P(t)|$ are the intensity fluctuations of the active and passive phases far from the interface.
Here,  $-1 \leq \phi \leq +1$, with -1 corresponding to the completely passive phase and 1 corresponding to the completely active phase. 

The interfacial region is fuzzy with finite thickness due to the intermingling of active and passive bacteria. Following classical phase-field approaches, we mathematically define the interface location as the locus of
points satisfying $\phi = 0$. 

{(II) From velocity fields through PIV:} The second order parameter $\phi_v$ is based on the velocity fields calculated from PIV (SI-$\S$III-2, SI-Fig. 3). We define 
\begin{equation}
\phi_v({\bf r},t)=\frac{2v^2({\bf r},t)-v_A^2-v_P^2}{v^2_A-v^2_P}
\end{equation}
where $v_A^2$ and $v_P^2$ are velocities of the active and passive phases far from the interface and ${\bf r}$ is the position of velocity vectors sampled at 3 $\mu$m intervals. Again we have the bounds $-1 \leq \phi_{v} \leq +1$ with 
the interface location defined as the locus of
points satisfying $\phi_{v}= 0$. 

\subsection{{Position and thickness of the diffuse interface}}
\label{M5}
\noindent{}The boundary separating the active and passive phases resembles a propagating and broadening interface. To analyze this behavior, we fit the order parameter profiles to the classical form for a quasi-equilibrium diffuse, moving interface. We use the order parameter fields extracted from intensity fluctuations (Fig. \ref{F2}) and PIV data (SI-$\S$III) to obtain averaged one-dimensional phase profiles. 

For experiments with the [H] aperture, we choose cartesian coordinates $(x,y)$ with the $y=0$ line aligned with the edge of the exposure. For times post-exposure, fields are averaged in the $x$ direction to obtain $\phi^{*}(y,t)$. Data from experiments then is fit to
\begin{equation}
\phi^{*} ={\mathrm{tanh}} \Bigg[{{y-d(t)}\over w(t)}\Bigg], \:\:\:\:\:\:{\mathrm{and}} \:\:\:\:\: \phi^{*}_{v} ={\mathrm{tanh}} \Bigg[{{y-d_{v}(t)} \over w_{v}(t)}\Bigg].
\end{equation}
Here $d(t)$ and $ d_{v}(t)$ denote the interface location and $w(t)$ and $w_{v}(t)$ denote the diffuse interface thickness. 

In experiments using the [O] exposure, we define a polar co-ordinate system with center at the center of the passive phase and obtain an azimuthally-averaged, radially-dependent value of $\phi^{*}(r,t)$. 
Reduced data is then fit to the forms 
\begin{equation}
\phi^{*}={\mathrm{tanh}} \Bigg[{{r-d(t)}\over w(t)}\Bigg], \:\:\:\:\:\:{\mathrm{and}} \:\:\:\:\: \phi^{*}_{v} ={\mathrm{tanh}} \Bigg[{{r-d_{v}(t)} \over w_{v}(t)}\Bigg]
\end{equation}
where now $d(t)$ is the averaged radius of the passive domain.
Since $r=0$ is the center of the passive region, equation (7) will exhibit significant errors for times close to dissolution ($d \leq w$). 

The length-scale over which  $\phi$  varies sharply, $w(t)$, yields the density correlation length characterizing  the penetration of active bacteria into the passive phase. The length-scale over which  $\phi_{v}$  varies sharply, $w_{v}(t)$, yields the momentum penetration length - the length over which passive bacteria are pushed around by the active bacteria without being completely dislodged. 

\noindent
\subsection{Interface-flow coupling}
\label{M6}

\noindent
\paragraph*{(i) Height fluctuations}
\noindent
To compare the structural features of the collective flow with the features of the interface, we calculated the static structure factor of the interface after suitably averaging fluctuations on the order of bacterial lengths. Focusing on experiments with the [H] aperture, we compute the height fluctuation 
$\Delta h(t,x)=h(t,x)-d(t)$, where $h(t,x)$ is the interface position interpolated as a function of the arc-length coordinate $x$ and $d(t)$ is the mean interface position. Next, the wavenumber dependent Fourier modes of the height fluctuations, $\Delta h_q (t)$, was obtained as 
\begin{equation}
\Delta h_q (t) = \frac{1}{\mathcal{L}_x} \int_{0}^{\mathcal{L}_x} \Delta h(x,t) \:\: e^{-iqx}\:\:dx.
\end{equation}
Here, the wave number is $q=n \pi/{\mathcal{L}}_x$ where $\mathcal{L}_x \approx 200$ $\mu$m). We varies $q$ from 0.015 $\mu$m$^{-1}$, which corresponds to the system size, to 1.0 $\mu$m$^{-1}$, which corresponds to approximately 3 $\mu$m {($\sim$ half a bacterial length)}. {The static structure factor was determined as the temporal-average of the Fourier mode magnitude square, $\langle |\Delta h_q|^2 \rangle$.}

\noindent{}The energy transfer to the surface comes from the flux of momentum flowing into the passive domain from the active domain as active bacteria invade and erode the interface. 
\noindent
\paragraph*{(ii) Energy spectra}
\noindent
The two-dimensional energy spectra of the flow is defined here through the PIV-velocity fields as
\begin{equation}
E(q)=\frac{q}{2\pi}\int d{\bf R} \: e^{-i {\bf q}\cdot {\bf{R}}}\langle {\bf v}(t,{\bf r}_{0})\cdot {\bf v} (t,{\bf r}_{0}+{\bf R})\rangle, 
\end{equation}
where the brackets here denote averages over ${\bf r}_{0}$.
For the near-interface case, we calculate $E(q)$ within an [80 $\times$ 80] $\mu$m$^2$ area, adjacent to the interface ($Y$ ranging from 0-80 $\mu$m; average distance 40 $\mu$m). For the far from interface case, we calculate $E(q)$ within an [100 $\times$ 100] $\mu$m$^2$ area, 100 $\mu$m$+$ from the interface ($Y$ ranging from 80-180 $\mu$m; average distance 130 $\mu$m). 
We varied $q$ from approximately 0.03 $\mu$m$^{-1}$ to 1.0 $\mu$m$^{-1}$, corresponding to approximately the size of the region (80 $\mu$m) and to {the bacterial velocity field spacing} (3 $\mu$m), respectively. For the [H]-aperture results (Fig. 3), the interface position is a flat propagating line, and  $Y$ is the distance normal to this line.

\paragraph*{(iii) Temporal correlation functions}
We calculated the autocorrelation function of the boundary position (using the [H]-aperture), defined here as
\begin{equation}
 C_{\textrm{int}} (\Delta t)  = \frac{\langle \Delta h(t_0) \Delta h(t_0+\Delta t)\rangle}{\langle |\Delta h(t_0)|^2\rangle},
\end{equation}
where the correlation is averaged over reference locations and reference times $t_0$. The data is fit to a single decaying exponential, yielding a characteristic decay time $\tau$ of 16~s ({SI-Fig. 9}). For characterize the bacterial flow, we measure the normalized spatial correlation of the velocity director {\bf v} (from PIV) as a function of time $\Delta t$ and the distance normal to the interface $Y$, as given by 
\begin{equation}
C_{\textrm{flow}} (\Delta t,Y) = \frac{\langle {\bf v}(t_0,Y) {\bf v}(t_0+\Delta t,Y)\rangle}{\langle  {\bf v}(t_0,Y)^2\rangle} 
\end{equation}
where the correlation is averaged over reference locations and reference times $t_0$. The flow correlations are also fit to a single decaying exponential to determine characteristic flow time scales as a function of $Y$ ({SI-Fig. 9}).

\subsection{{Interfacial stiffness and effective temperature}}
\label{M7}

\noindent{}To extract an effective interfacial stiffness  $\kappa$, we fit the structure factor in Fig. \ref{F3}b to $({\kappa A q^2})^{-1}$, assuming that the interfacial area is $h\times L_X$, where $h=1$ $\mu$m is the cell width (since the swarm is approximately a monolayer thick \cite{Darnton2010}) and $L_X$ is the length of the observed interface, 200 $\mu$m. 

To estimate an effective temperature of the swarm, we use two micron polystyrene spheres as probes of the swarming flows in the active region of the colony (SI-Fig. 10). The polystyrene particles are cleaned by centrifugation and then suspended in a buffer solution (67 mM NaCl aqueous solution) with a small amount of surfactant (Tween 20, 0.03\% by volume). A small aliquot of this particle solution (20 $\mu$L) is gently pipetted unto  the bacterial colony in a location where expanding colony front meets the agar. After the polystyrene particles are introduced, we allow the colony to settle for 5 minutes before imaging. We do not observe any change in the behavior of the swarming bacteria due to the addition of these particles at 0.8 \% area fraction.
We track the particle positions over time using standard particle tracking techniques. We define the (time-averaged) particle speed as the particles displacement (in two dimensions) over a 1 second time interval, allowing tracers to sample multiple vortex structures (characteristic lifetimes $\sim$ 0.1 second).  
The particle speed distributions seem to follow a 2D Maxwell-Boltmann distribution  (SI-Fig. 10b), $p(v) = v _{m} (k_BT_{\textrm{eff}})^{-1} \textrm{exp}(-mv^2/2k_BT_{\textrm{eff}})$, where $m$ is the mass of the polystyrene particle, $k_B$ is the Boltzmann constant, and $T_{\textrm{eff}}
\approx 2.2 \times 10^{5}$ K, approximately 700 times the thermal temperature (293 K). This effective temperature is to be interpreted as a mixture temperature purely due to the energy in the swarming collective flows.

\subsection{{Effects of curvature and flow on interface speed}}
\label{M8}
\noindent{}Does an approximate linear relationship between the interface speed, curvature, and a characteristic of the bacterial flow exist for small propagation speeds? Such a relationship may be viewed then as an {\em active} analogue of the classical Gibbs-Thomson-Stefan boundary conditions for passive systems.  Swarming provides the impetus for the motion and keeps the active-passive interface always out of equilibrium. We can directly gauge the intensity of the flow and its activity via  measurements such as PIV or particle tracking methods. This is preferable to using temperature or chemical potential that cannot be measured directly and exact forms of which are a matter of debate \cite{Takatori2014,Cates2015,Patteson2016_2}. 

To test for correlations between the interface speed, curvature and flow speeds we assume that two independent processes act on the interface leading to erosion and remodeling.  The first arises from self-propulsion based interactions at the single bacteria level and is proportional to density of active bacteria $\rho_{\mathrm{A}}$. The second acts at longer spatiotemporal scales and accounts for hydrodynamic interactions and collective flow. 
Let ${v}_{\mathrm{int}}({\bf r})$ be the interface velocity at location ${\bf r}$ on the interface; the (mesoscale) normal at this location 
is ${\bf n}({\bf r}, t)$, the (mesoscale) tangent  is ${\bf t}({\bf r}, t)$, and the (mesoscale, collective) bacterial velocity field is ${\bf v}$. Let ${\bf Q}$ be the alignment tensor that quantifies the orientation of the jammed passive bacteria in the neighborhood of ${\bf r}$. 
Ignoring quadratic and higher order terms (in $\bnab {\bf n}$, ${\bnab {\bf {t}}}$,
$\bnab {\bf v}$ and ${\bnab \bf {Q}}$), we write the general linear form
\[
{v}_{\mathrm{int}} {\bf n} \:\: \approx \:\: \rho_{\mathrm{A}} {\bf A}_{1} +  \rho_{\mathrm{A}} {\bf A}_{2}\: ( {\bnab \bcdot {\bf n}} ) + 
 \rho_{\mathrm{A}} ({\bf n}\bcdot{\bf Q}\bcdot{\bf n})\: {\bf v}\: +  \]
 \[
 \rho_{\mathrm{A}}({\bf n}\bcdot{\bf Q}\bcdot{\bf t})\: {\bf v} +  \rho_{\mathrm{A}} ( {\bf t}\bcdot{\bf Q}\bcdot{\bf t} )\: {\bf v} +  \rho_{\mathrm{A}} ({\bf t}\bcdot{\bf Q}\bcdot{\bf n}) \:{\bf v}\: +
 \]
 \begin{equation}
 \rho_{\mathrm{A}} ({\bf v}\bcdot{\bf Q}\bcdot{\bf n} + {\bf n}\bcdot{\bf Q}\bcdot{\bf v} + {\bf t}\bcdot{\bf Q}\bcdot{\bf v} + {\bf v}\bcdot{\bf Q}\bcdot{\bf t}) 
\end{equation}
with  ${\bf A}_{1}$ and ${\bf A}_{2}$ being functionals of ${\bf Q}$, ${\bf n}$ and ${\bf t}$. 
The first term on the right hand side reflects {\em local} density driven self-propulsion effects present even in the absence of curvature or collective motion. The second curvature term reflects variations in the erosion rate due to geometric effects and bacteria preference to reside near the surface. The other terms describe erosion driven by the collective flow and include normal and tangential velocity contributions. 
Equation (12) as written cannot be used directly to interpret our data as we do not measure or visualize the alignment field of bacteria in the passive region. 

To gain insight into the form of this linear relationship, we studied the trajectories of 2 $\mu$m polystyrene tracer particles trapped in the passive phase to identify the sequence of mechanistic events leading to their eventual extraction (Fig.\ref{F4}a). From observations on dozens of tracers, we find that erosion occurs due to initially trapped passive particles being dislodged from their neighbors, then moving parallel to the surface as they are sheared by active bacteria (SI-Movie 6), and eventually escaping and leaving that passive domain due to normal streaming flows that occur between vortices at the interface  (Fig.\ref{F4}a). 
Guided by these observations of tracer particles escaping from the passive phase (Fig.\ref{F4}a), we further deduce that anisotropic caging effects result in tangential and normal swarm velocities to affect the dislodgment of the passive particles differently.  At the simplest level this incorporates the non-isotropic nature of ${\bf Q}$.
Incorporating these ideas, we write 
 \begin{equation}
{v}_{\mathrm{int}} \approx    \rho_{\mathrm {A}} (\alpha_{1} + \alpha_{2}\:{\mathcal{C}} +
\alpha_{3} ({\bf v}{\bcdot} {\bf n}) + \alpha_{4} ({\bf v}{\bcdot} {\bf t})). 
\end{equation}
Deviations from this form can be attributed to non-linear effects involving curvature and flow, neglect of alignment effects in the passive phase and density variations in the active phase. While we expect $\alpha_{1}$ (and thus $a_{1} = \rho_{\mathrm {A}} \alpha_{1}$) to be zero in the absence of flow and curvature, we retain this as a fitting constant as higher order terms and the colony front velocity may lead to a net interface propagation.

\section*{Acknowledgements}
\noindent
We thank Edward Steager, Elizabeth Hunter, Somayeh Fahardi, Mark Goulian, Prashant Purohit, Julia Yeomans, and James Sethna for fruitful discussions. This work was supported by NSF-CBET-1437482. A. E. Patteson was supported by an NSF Graduate Research Fellowship.

\section*{Author Information}
\noindent
The authors declare no competing financial interests. \\
$^*$Correspondence and requests for materials should be addressed to A.E.P. (aepattes@syr.edu) and P.E.A. (parratia@seas.upenn.edu).

\end{document}